\documentclass[letterpaper,twocolumn,10pt]{article}
\usepackage{usenix2019_v3}

\usepackage{tikz}
\usepackage{amsmath}

\usepackage{comment}
\usepackage{booktabs} % For formal tables
\usepackage{xspace}
\usepackage[normalem]{ulem} % For underlining

\usepackage{algorithm}
\usepackage{algpseudocode}

\usepackage{subcaption}
\usepackage{color, colortbl}
\definecolor{TableRowColor}{rgb}{0.88,1,1}

\newcommand{\opt}{{\tt 3DXP}\xspace}

\newcommand{\metric}{{\tt TRaCaR}\xspace}

\usepackage{etoolbox}
\makeatletter
\patchcmd{\maketitle}
	{\@maketitle}
	{\@maketitle\vspace{-1.5em}} % Change this as needed
	{}
	{}
\makeatother

\usepackage{filecontents}

\begin{document}
%-------------------------------------------------------------------------------

%don't want date printed
\date{}

\title{The \metric Ratio: Selecting the Right Storage Technology for Active Dataset-Serving Databases}

\author{
{\rm Francisco Romero}\\
Stanford University
\and
{\rm Benjamin Braun}\\
Stanford University
\and
{\rm David Cheriton}\\
Stanford University
% copy the following lines to add more authors
% \and
% {\rm Name}\\
%Name Institution
} % end author

\maketitle

%-------------------------------------------------------------------------------
\begin{abstract}
%-------------------------------------------------------------------------------
Main memory database systems aim to provide users with low latency and high throughput access to data.
Most data resides in secondary storage, which is limited by the access speed of the technology.
For hot content, data resides in DRAM, which has become increasingly expensive as datasets grow in size and access demand.
With the emergence of low-latency storage solutions such as Flash and Intel's 3D XPoint (\opt), there is an opportunity for these systems to give users high Quality-of-Service while reducing the cost for providers.

To achieve high performance, providers must provision the server hosts for these datasets with the proper amount of DRAM and secondary storage, as well as selecting a storage technology.
The growth of capacity and transaction load over time makes it expensive to flip back-and-forth between different storage technologies and memory-storage combinations.
Servers set up for one storage technology must now be reconfigured, repartitioned, and potentially replaced altogether.
%Furthermore, it is unacceptable for providers to under-provision their resources.
As more low-latency storage solutions become available, how does one decide on the right memory-storage combination, as well as selecting a storage technology, given a predicted trend in dataset growth and offered load?

In this paper, we describe and make the case for using the \metric ratio --- the transaction rate divided by the storage capacity needed for a workload --- for allowing providers to choose the most cost-effective memory-storage combination and storage technology given their predicted dataset trend and load requirement.
We explore how the \metric ratio can be used with \opt and Flash with a highly-zipfian b-tree database, and discuss potential research directions that can leverage the ratio.
\end{abstract}

%-------------------------------------------------------------------------------
\section{Introduction}
%-------------------------------------------------------------------------------
\begin{figure}[t!]
    \centering
    \includegraphics[scale=0.50]{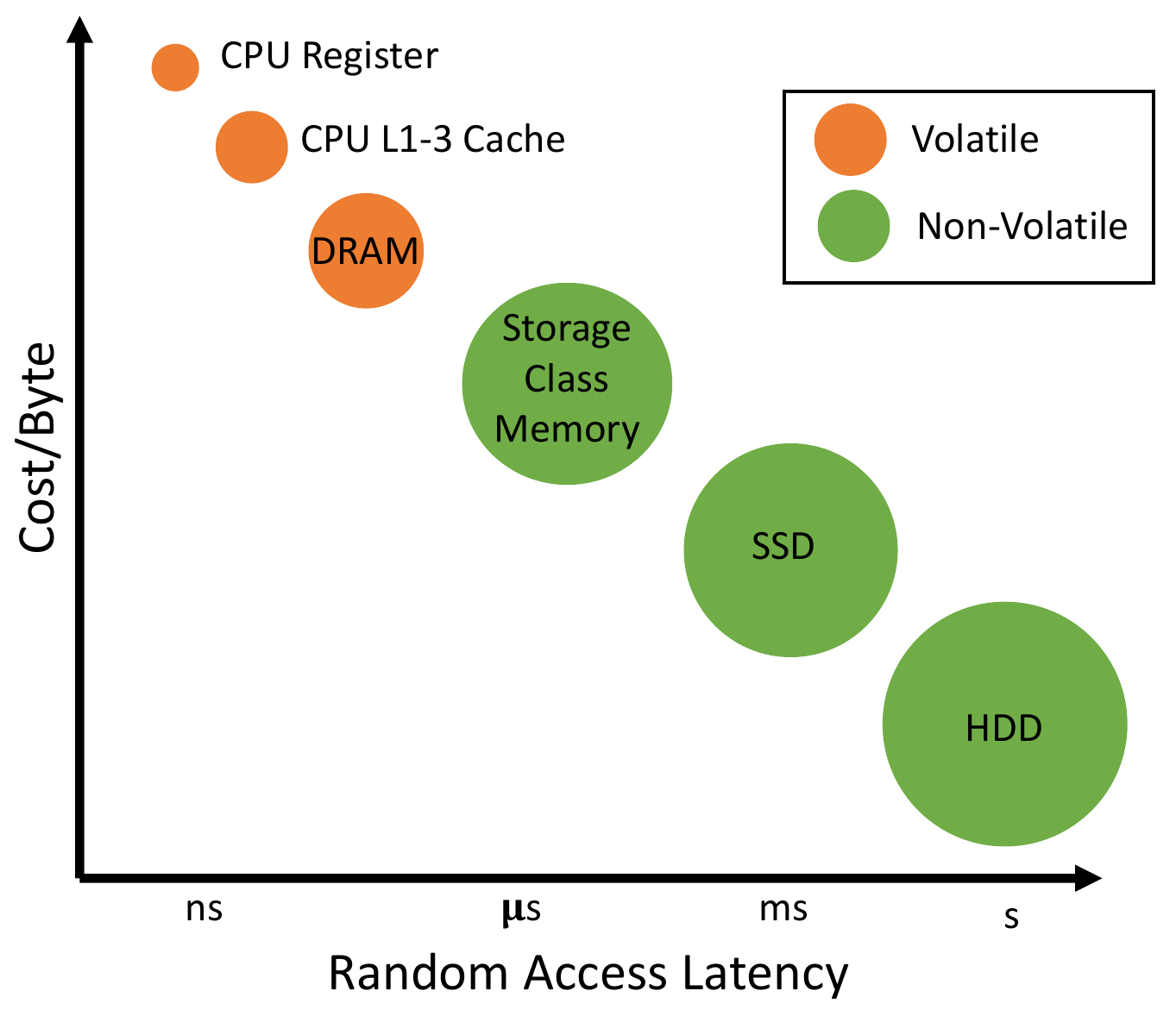}
    \caption{\small Comparing the cost and random access latency of storage technologies~\cite{WestDig}. Each technologies diameter approximates the capacity relative to the other technologies. \opt is an example of a Storage Class Memory.}
    \label{fig:memhier}
\end{figure}

Datasets actively used by applications today are growing in size and access demand, making it difficult for providers such as Amazon and Google to keep transaction latencies low and throughput high.
A provider is tasked with finding the right balance between the amount of DRAM needed to service hot content actively being operated on, and secondary storage for keeping the majority of the dataset.
However, over the last few years, tight supplies from DRAM vendors has led to increases in DRAM cost. For example, from 2016-2017, DDR4 increased in cost by 2.3$\times$~\cite{Eisenman:2018:RDF:3190508.3190524,EPSNews}.
Hence, keeping all content in DRAM, or over-provisioning the amount of DRAM needed to service the active content of one's dataset can lead to a high total cost of ownership (TCO).
Furthermore, when transactions cannot be serviced from main memory, performance degrades~\cite{Oukid:2016:FHS:2882903.2915251,Stoica:2013:EEO:2485278.2485285}.
%In these situations, the operating system (OS) is tasked with writing out cold pages to secondary storage, and when an evicted page is accessed, the OS will read the page back in via a page fault.
%During this time, the database transaction will stall, and noticeable performance loss is experienced.

The increasing size and access demand of datasets has led to newly proposed storage technology solutions called \emph{Storage Class Memories}. These are characterized by being non-volatile, having short access times, and having lower cost-per-bit compared to DRAM. Figure~\ref{fig:memhier} shows that Storage Class Memories lie between SSD (Flash) and DRAM in terms of both cost and access time, making such technologies viable for servicing page faults on the order of microseconds. For example, in-production today is Intel's 3D XPoint (\opt). With access latencies of about 10$\mu$s, \opt has an order of magnitude higher performance and durability than NAND Flash~\cite{Optane}. Thus, the technology offers a promising solution for databases that require low-latency access to secondary storage. However, this performance comes at a cost: \opt is currently about 6$\times$ more expensive than Flash~\cite{neggFlash,neggOpt}.
Therefore, in addition to configuring the proper amount of DRAM and secondary storage, a provider would need to decide on whether a more expensive, low-latency storage technology is necessary to meet their access demands.

Choosing the right storage backend for a provider's dataset has performance \textit{and} cost implications.
For example, previous work has shown that an SQL equijoin query on two tables of over 100GB in size can have a performance variation of over 30$\times$ and a cost difference of 8$\times$ when selecting between storage technologies~\cite{Klimovic}.
Providers will typically over-provision the amount of main memory and storage needed, or select the most expensive secondary storage technology (e.g., choosing \opt over Flash).
Under-provisioning the datacenter is unacceptable, as strict Quality-of-Service can be violated.

The growth in capacity and offered load makes it expensive, both in cost and time, to constantly have to revisit these decisions and flip back-and-forth between different storage technologies, as well as main memory-secondary storage setups.
Providers would need to migrate data, optimize transactions for the underlying storage technology, and potentially switch out the servers themselves if the CPU, DRAM, or slots (e.g., PCIe) are insufficient.
Given a predicted performance and capacity growth \emph{trend}, providers need a way to make these decisions to save cost and avoid constant reprovisioning.
However, existing studies~\cite{CostPerformanceDataStores,Ousterhout:2010:CRS:1713254.1713276,NarayananEurosys} that attempt to define which storage technology is most suitable for a given workload (1) consider technology such as DRAM, Flash, and Hard Disk separately (i.e., it all resides in only one), (2) do not consider Storage Class Memory technologies such as \opt, and (3) do not consider the expected \emph{trend} of one's dataset growth.

In this paper, we introduce the \uline{T}ransaction \uline{Ra}te - \uline{Ca}pacity \uline{R}equirement --- \texttt{TRaCaR} --- ratio (pronounced ``tracker'' ratio). The \metric ratio gives providers a way to both select a storage technology and configure main memory-storage capacity based on the expected trend between offered transaction load and dataset capacity growth. We make that case for the \metric ratio's efficacy by considering whether to provision between \opt and Flash for a database with a highly-zipfian access pattern with different read/write ratios.
In doing so, we show how a provider would use the computed \metric given their growth trend.
%Finally, we explore open research questions and future directions related to the \metric ratio.

\begin{comment}
The rest of this paper is organized as follows. Section~\ref{sec:eval} demonstrates how the \metric ratio is used by a provider. Section~\ref{sec:whyrevisit} discusses our rationale and the type of datasets we target. Section~\ref{sec:comptracar} shows how the \metric ratio is computed. Section~\ref{sec:sensitivity} analyzes \metric's sensitivity to cost. Finally, Section~\ref{sec:relwork} overviews related work, and Section~\ref{sec:conclusion} summarizes and concludes this study.
\end{comment}

\begin{figure}[t!]
    \centering
    \includegraphics[scale=0.47]{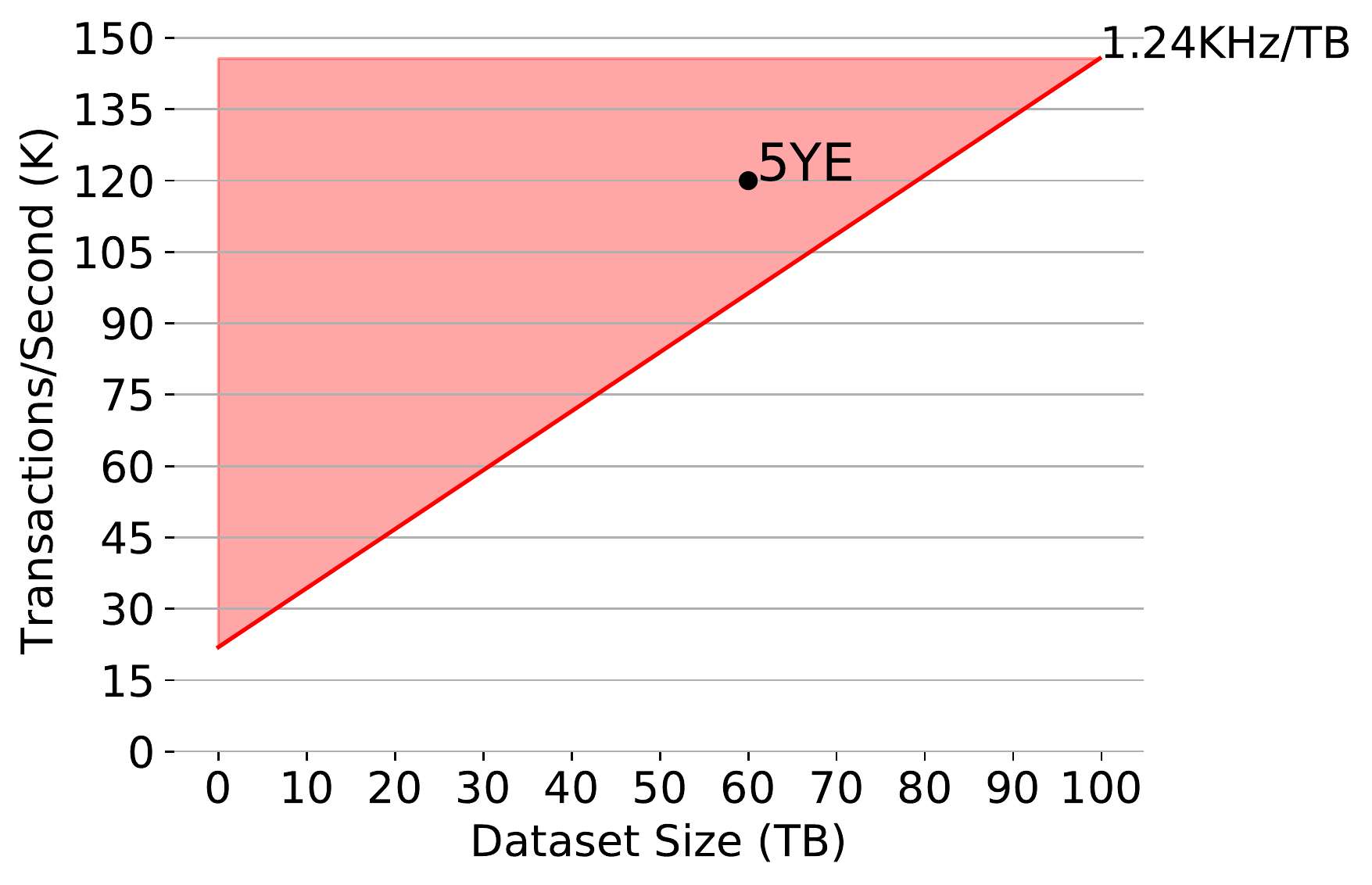}
    \caption{\small The \metric ratio for when \opt (shaded region) is more cost-effective versus Flash for 50\% random reads and writes.
    \texttt{5YE} denotes the 5-year expected requirements from Section~\ref{sec:confserv}.}
    \label{fig:half_reads}
\end{figure}

%-------------------------------------------------------------------------------
\section{Using the {\large \metric} Ratio} \label{sec:eval}
%-------------------------------------------------------------------------------
To understand how providers would use the \metric ratio to provision their servers, we present a motivating example for a highly-zipfian database with 50\% random reads and writes.

In our example scenario, the provider starts with a 50\% random read/write dataset that is 1TB in size and has an access demand of 20,000 transactions per second.
The provider studies their dataset's growth in terms of both size and access demand over one year.
Over this period of time, the provider finds that their dataset has grown by 10TB, and their access demand has increased by 20,000 transactions per second.
Hence, the \metric ratio for their dataset is \textbf{2KHz/TB}, or the throughput has grown by 2,000 transactions for each TB of data.
The provider now wants to decide (1) how to select the host server setup
and (2) the storage technology that will minimize cost if their performance continues to grow according to the 2KHz/TB \metric ratio.

\begin{table}[t!]
  \centering
  %\small
  \begin{tabular}{p{4cm} p{3.5cm}}\hline
  Random Read/Write Mix    & \metric                                                              \\ \hline
  \textbf{100\% Reads}    & 1.25KHz/TB
                            \\ \hline
  \textbf{50\% Reads/Writes} & 1.24KHz/TB                                                         \\ \hline
  \textbf{100\% Writes}      & 1.17KHz/TB                                                         \\ \hline
  \end{tabular}
  \caption{\small The \metric ratio for selecting between \opt and Flash for different random read/write mixes.
  \opt is cost-optimal if a provider's predicted trend \metric is greater than the \metric listed in a given row.
  }
  \label{tab:all_tracar}
\end{table}

\begin{table*}[t!]
  \centering
  %\small
  \begin{tabular}{lll}\hline
  Item & \opt Setup Cost (\textbf{3} Servers Needed)    & Flash Setup Cost (\textbf{19} Servers Needed)                                                               \\ \hline
  \textbf{DRAM} & \$4,950 & \$62,700                                                       \\ \hline
  \textbf{Secondary Storage} & \$72,000 & \$12,000                                                       \\ \hline
  \textbf{Processor} & \$1,200 & \$7,600                                                       \\ \hline
  \textbf{Miscellaneous} & \$3,000 & \$19,000                                                       \\ \hline
  \rowcolor{TableRowColor}
   \textbf{TOTAL} & \$81,150 & \$101,300                                                       \\ \hline
  \end{tabular}
  \caption{\small Comparing optimal cost setups between \opt and Flash for a dataset size of 60TB and a throughput requirement of 120,000 transactions per second.
  \opt is almost 25\% cheaper and requires less servers to meet the access needs.
  }
  \label{tab:sample_cost}
\end{table*}

\subsection{Selecting a Storage Technology} \label{sec:selstore}
Figure~\ref{fig:half_reads} shows the \metric ratio for when \opt is more cost-effective compared to Flash for 50\% random reads and writes.
\opt's cost-effective region is shaded, and its corresponding \metric ratio is shown over the breakpoint line.
Providers whose datasets \metric ratio are larger than \\1.24KHz/TB (such as in our motivating example) should select \opt for their secondary storage technology, as it is more cost-effective for their performance trend.
Similarly, if a provider's \metric ratio is smaller than the breakpoint shown in Figure~\ref{fig:half_reads}, Flash should be selected.

Table~\ref{tab:all_tracar} shows the \metric ratio for other random read and write mixes.
Providers can compare their dataset's workload to the corresponding \metric ratio as described above to determine which storage technology is most cost-effective given their performance trend.
If the provider's \metric is greater than the value in Table~\ref{tab:all_tracar} for a given random read/write mix, \opt is cost-optimal.
For our motivating example, \opt would still be cost-optimal for different read/write mixes.

\subsection{Configuring the Servers} \label{sec:confserv}
After a provider uses their computed \metric ratio to select the most cost-effective storage technology, they must provision the datacenter with the right number of servers, as well as amount of DRAM and secondary storage in each server.
To do this, the provider finds the point along their \metric ratio that represents their predicted capacity and access demand at a future time (e.g., 5-10 years out), and computes the lowest cost-setup for the corresponding secondary storage technology.
In our motivating example, if we are looking 5 years into the future (denoted as \texttt{5YE}, or 5-year expected in Figure~\ref{fig:half_reads}), this would correspond to a dataset size of 60TB and 120,000 transactions per second.

Table~\ref{tab:sample_cost} shows the optimal cost setup using \opt to meet this requirement, and for reference, also shows the Flash setup.
For our motivating example, the predicted trend's setup is almost 25\% cheaper by using \opt over Flash, since \opt requires fewer servers and less DRAM.
A provider's \metric ratio can also be used to determine how many new servers to provision per year to meet the performance demands.
Further details about the cost model used are described in Section~\ref{sec:costs}.

Determining the lowest cost setup for a given storage technology and discussing how the \metric ratio of Figure~\ref{fig:half_reads} was computed is detailed in Section~\ref{sec:comptracar}.

\subsection{Variations in Predicted Trend} \label{sec:extracost}
Determining how one's dataset will grow 5-10 years into the future can be difficult.
In addition, secondary storage and memory prices can vary over time.
To account for this, providers are often willing to accept a maximum cost difference if it means not having to switch between different storage technologies.
For example, one may be willing to pay up to 10\% more for an \opt setup if it decreases the \metric ratio in \opt's favor.
This corresponds to computing the \metric ratio in Figure~\ref{fig:half_reads} for setups in which \opt is at most $E$\% more expensive, where $E$ is the \emph{extra cost} that providers are willing to pay.
Providers can then select a storage technology and server setup as described in Sections~\ref{sec:selstore} and~\ref{sec:confserv}. A cost sensitivity analysis for \metric is discussed in Section~\ref{sec:sensitivity}.

%-------------------------------------------------------------------------------
\section{Why {\large \metric}} \label{sec:whyrevisit}
The \metric ratio is motivated by the need to configure host servers for high throughput, low latency database systems as the capacity and access demands change over time.
The following observations guided us in our development and analysis of the \metric ratio:
\begin{itemize}
    \item \textbf{O1:} Devices such as \opt have latencies that are low enough such that one can focus on the throughput needs for a particular dataset.
    \item \textbf{O2:} Selecting between storage technologies need not be mutually exclusive (e.g., not all the data needs to reside in DRAM). A combination of DRAM and secondary storage can be significantly cheaper than only using DRAM. Furthermore, one can still benefit from storing the hot content being operated on in DRAM.
    \item \textbf{O3:} Content that would normally be completely stored in DRAM tends to have a zipfian access pattern, where a small percentage of the data accounts for the majority of accesses.
\end{itemize}
To the best of our knowledge, there is no existing work in evaluating storage cost-performance that captures all three of these observations in their analysis. 

\subsection{Target Dataset} \label{sec:inttargexp}
Based on \textbf{O3}, we envision using \metric for datasets that could be split across DRAM and low-latency secondary storage backends. To clarify what these datasets look like, we first describe the two extreme dataset types before introducing \emph{Active Datasets}.

\noindent{\bf{Extreme Latency Sensitivity. }}Workloads with sub-10$\mu$s latency requirements cannot tolerate the latency of going to secondary storage --- even with a low-latency technology such as \opt. Thus, they must reside in DRAM, and the decision of what storage backend to use becomes irrelevant.

\noindent{\bf{Archival Data. }}Rarely-accessed data with low throughput requirements is normally stored in low-cost storage facilities. For example, Facebook has built cold-storage facilities that can hold up to 1,000PB (one exabyte) of ``less popular'' data and replicas~\cite{FBCold}. These large and infrequently accessed datasets are best served using inexpensive disks, and are not discussed in this paper~\cite{Andersen:2009:FFA:1629575.1629577}.

\noindent{\bf{Active Datasets. }}The focus for \metric is on Active Datasets, which have the following characteristics. First, they are on the order of hundreds of GBs to several tens of TBs. Second, their data is frequently accessed and exhibits a zipfian access pattern. Third, they have high throughput and low latency requirements. Examples of such datasets include popular stocks in high-frequency trading or popular songs on music streaming platforms.
Today's Active Datasets are almost entirely stored in DRAM to meet their stringent performance requirements.
While DRAM offers low latency and high throughput benefits, it has drawbacks:
\begin{itemize}
    \item Storing hundreds of GBs to several tens of TBs in memory is expensive, both in terms of the DRAM itself and the servers needed to service the dataset.
    \item DRAM has reliability limitations: if there is a sudden loss of power or system malfunction, recovering lost content from DRAM can take hours, especially if the amount of data lost is on the order of TBs.
\end{itemize}

\subsection{Summary}
From these observations, the challenge lies in finding the most cost-effective combination of DRAM-secondary storage server setup for a predicted performance trend. The \metric ratio encompasses these observations and allows a service provider to characterize their dataset in terms of size and offered load to decide what secondary storage technology and server setup is cost-optimal over the next 5-10 years.

%-------------------------------------------------------------------------------
\section{Computing {\large \metric}} \label{sec:comptracar}
%-------------------------------------------------------------------------------
\begin{algorithm}[t!]
  \begin{algorithmic}[1]
        \Function{ComputeTRaCaR}{\texttt db,workld,stor1,stor2}
        \State $\texttt{tpMem1 = getThroughputMem(db,workld,stor1)}$
        \State $\texttt{tpMem2 = getThroughputMem(db,workld,stor2)}$

        \State $\texttt{validSetup1 = getSetups(stor1,tpMem1)}$
        \State $\texttt{validSetup2 = getSetups(stor2,tpMem2)}$

        \State $\texttt{tracar = getTRaCaR(validSetup1,validSetup2)}$
        \State $\texttt{\textbf{return} tracar}$
        \EndFunction

  \end{algorithmic}
  \caption{\small Computing \metric}
  \label{alg:costmodel}
\end{algorithm}

In this section, we describe how we compute the \metric ratio used in our motivating example throughout Section~\ref{sec:eval}.
We focus on \opt and Flash for one particular database, but this methodology generalizes to any Active Dataset-serving database, set of storage backend technologies, and storage hierarchies (e.g., an \opt tier backed by Flash).

\subsection{Methodology} \label{sec:costmodel}
Our methodology for computing \metric for low-latency storage technologies is based on observations \textbf{O1} and \textbf{O2} in Section~\ref{sec:whyrevisit}.
The objective is to find the cost break point (i.e., the \metric ratio) between the secondary storage technologies under comparison given a range of throughput and dataset size requirements.
For each \textit{(dataset size, throughput)} tuple, we want to find the lowest cost setup.

Algorithm~\ref{alg:costmodel} describes how we compute the \metric ratio.
\texttt{db} is the user's database, \texttt{workld} is the workload (e.g., 50\% random reads/writes), and \texttt{stor1}, \texttt{stor2} are the storage technologies under consideration (e.g., \opt and Flash).
Given the workload, \texttt{getThroughputMem} computes the throughput for the database as the amount of DRAM is reduced. This allows us to create a throughput versus memory reduction percentage relationship for a particular storage technology.
After computing this for both technologies, we get all the valid server setups for each of the storage technologies with \texttt{getSetups}.
A valid server setup entails figuring out whether a given amount of DRAM, storage devices, and nodes can meet the throughput requirement for an inputted dataset size.
We use the output of \texttt{getThroughputMem} to determine how much DRAM is necessary for meeting a given throughput and dataset size requirement.
Finally, \texttt{getTRaCaR} finds (1) the cheapest storage technology and (2) the cheapest server setup for all \textit{(dataset size, throughput)} pairs.
The \metric ratio is then determined by finding the breakpoint line separating one storage technology from being more cost-effective over the other.
As done in Section~\ref{sec:eval}, providers can then compare their predicted performance trend against the \metric ratio computed for the storage technologies under consideration.

In \texttt{getSetups}, to increase the throughput of our system, we add more nodes, and spread the dataset across the nodes \cite{SmallCacheBigEffect}. Thus, as the number of nodes increases, the amount of storage \emph{per node} decreases, but the total amount of storage stays the same. For instance, to double the throughput of a single node system, we would add a second node, and put half of the dataset in each node. We assume storage device costs are linear in their capacity when computing the cost of a complete system setup.

Depending on the database and workload being profiled, computing the \metric ratio can take tens of minutes (i.e., the time taken in \texttt{getThroughputMem}). \texttt{getSetups} and \texttt{getTRaCaR} are both currently greedy configuration searches implemented using techniques from~\cite{videostorm}. However, we envision providers only needing to do this \emph{once} when initially provisioning their servers.

%While having a representative model for computing \metric is important for standardizing how providers select a storage technology and server setup given their performance trend, additions or considerations to be added to our proposed \texttt{computeTRaCaR} algorithm are discussed in Section~\ref{sec:limfutwork}.

\subsubsection{Cost Model} \label{sec:costs}
To compute the cost of a setup, we use traditional server nodes that use \$400 processors and server equipment and maintenance costs of \$1,000~\cite{GRCooling,Andersen:2009:FFA:1629575.1629577}. At the time of writing, Flash is about \$0.20/GB~\cite{neggFlash}, \opt is about \$1.20/GB~\cite{neggOpt}, and DRAM is about \$5.50/GB~\cite{neggDRAM}.

\subsection{Abstracting the Database} \label{sec:bmark}
We abstract a database servicing Active Datasets with a highly-zipfian b-tree benchmark. The b-tree benchmark has the following attributes:
\begin{itemize}
    \item We performed 1 million random transactions on a key-value dataset. We evaluated three different read/ write mixes: 100\% reads, 50\% read/writes, and 100\% writes.
    \item B-tree nodes are multiples of the page size, and are page-aligned. This also means that accesses to secondary storage are page size multiples.
    \item Transactions have a zipfian distribution such that about 80\% of the accesses go to 20\% of the b-tree's values. Thus, 20\% of the b-tree's values are the hot-content.
\end{itemize}

\subsection{Configuring the Storage Backends} \label{sec:confstor}
All b-tree memory accesses are memory-mapped to Flash and \opt partitions. For Flash, we use Intel's SSD DC 3600~\cite{ssd3600} and for \opt we use Intel's Optane SSD DC P4800X \cite{ssd4800}.
Both devices are connected over PCIe NVMe 3.0x4 and have an HHHL Form Factor.
To vary the amount of memory available to the b-tree application as a function of the data size, we used Linux CGroups~\cite{CGroups}.
In all of our experiments, we use a sufficient number of threads to maintain a high throughput and amortize the cost of a page fault (i.e., a sufficient queue depth).
For \opt, we found that 4 threads was sufficient, while for Flash, we used 16 threads.
We ran all experiments on a dual-socket Intel Xeon E5-2630 @ 2.30GHz (12 virtual cores per socket) and 64GB of memory. We used Ubuntu 16.04 and Linux kernel 4.4.0-116.

%-------------------------------------------------------------------------------
\section{{\large \metric} Sensitivity} \label{sec:sensitivity}
%-------------------------------------------------------------------------------
\metric's breakpoints are sensitive to the parameter costs described in Section~\ref{sec:costmodel}. For example, a reduction in cost per byte of Flash would increase the \metric ratio, while a reduction in cost per byte of \opt would have the opposite effect. We note that if DRAM were to increase in price, it would likely benefit \opt. To meet the throughput requirement, systems configured with a Flash backend require more servers than those with an \opt backend. In addition, we found that less than 5\% of the dataset needs to be in memory \emph{regardless} of the backend, thus making the additional servers less cost-effective for Flash. It should also be noted that, in the particular case of \opt, Intel's multi-billion dollar investment in the technology makes it unlikely that there will be significant variations in the cost/GB of the storage technology, or in the \metric breakpoint presented in Section~\ref{sec:eval}. However, depending on the storage technology, ``cost-volatility'' may be more significant, thus motivating the use of the extra cost \metric ratio described in Section~\ref{sec:extracost}.

%-------------------------------------------------------------------------------
\section{Related Work} \label{sec:relwork}
%-------------------------------------------------------------------------------
%\noindent{\bf{Cost-Performance Analysis. }}
Throughput versus dataset tradeoffs have been previously studied for Flash, DRAM, and Hard Disk technologies. The work most similar to ours is Narayanan \textit{et al.} that analyzed when and how Flash devices should be used over Hard Disk~\cite{NarayananEurosys}. For their workloads, they conclude that Flash would have to be 3-3,000 times higher in terms of capacity/\$ to be more cost-effective than Hard Disks. Our work differs in several aspects: First, we consider Active Datasets with highly-zipfian distributions. Second, we consider the use of DRAM with the secondary storage technology, whereas they only use the secondary storage (or use Flash as a caching layer). Third, our results show that low-latency technology, such as \opt can potentially be more cost-effective than Flash given a dataset and throughput \emph{trend}. RAMCloud also studied the throughput-versus-dataset tradeoff to justify using DRAM over Flash~\cite{Ousterhout:2010:CRS:1713254.1713276}. Lomet studied the cost versus performance for Deuteronomy and MassTree, and focused on particular techniques to improve the cost and performance of data caching systems~\cite{CostPerformanceDataStores}. FAWN~\cite{Andersen:2009:FFA:1629575.1629577} highlighted the need for multiple nodes in a cluster for both storage space and query rates, and demonstrated how different storage technologies combined with Fawn can affect a service provider's TCO. \metric is based on computing the most cost-effective number of servers and amount of DRAM for each secondary storage technology under consideration. In addition, the \metric ratio can be applied to any secondary storage technology.

\section{Conclusion} \label{sec:conclusion}
%-------------------------------------------------------------------------------
In this paper, we introduced and made the case for the \metric ratio: a promising technique to select a storage technology and provision host database servers based on the capacity and throughput \emph{trends} for Active Datasets. We showed how providers would use \metric to select between \opt and Flash for workloads with different read and write mixes.
Our findings show that Active Datasets, which are typically serviced from DRAM, can leverage low-latency devices such as \opt to save on cost while still meeting performance objectives.
As database datasets continue to grow, we hope this study will change how service providers provision their servers with their dataset trends in mind, and how researchers approach and quantify future database systems with low-latency backends.

%-------------------------------------------------------------------------------
\bibliographystyle{plain}
\bibliography{\jobname}

%%%%%%%%%%%%%%%%%%%%%%%%%%%%%%%%%%%%%%%%%%%%%%%%%%%%%%%%%%%%%%%%%%%%%%%%%%%%%%%%
\end{document}